\documentclass[prc,letterpaper,showpacs,12pt,floatfix,preprint]{revtex4}
\usepackage{amsmath}
\usepackage{amsfonts}
\usepackage{bm}
\usepackage{graphicx}

\begin{document}

\title{Augmenting the Gauge-Gravity Correspondence to include Hadron
Polarizabilities}
\author{Pervez Hoodbhoy }
\affiliation{Department of Physics\\
Forman Christian College\\
Lahore, Pakistan. \\
}

\begin{abstract}
ADS/CFT models have achieved considerable success in describing hadronic
properties such as masses and Regge trajectories. Even if the minimal vertex
that couples photons to structureless spin-zero fields is used, one still
ends up with electromagnetic form factors of hadrons that are in fair to
good agreement with experiment. However, contradicting both experiment and
intuition, this minimal model gives zero for hadronic electric and magnetic
polarizabilities. We show here that if effective vertices are used, and
axial and vector mesons are allowed to propagate as intermediate states,
then the static polarizabilities can in principle be computed from ADS/CFT.
\end{abstract}

\maketitle



Following the Maldacena revolution \cite{Theory}, hundreds of papers have
been written over the last two decades in attempts to build a theory of
hadronic structure based upon the correspondence between Anti
de-Sitter/conformal field theory (AdS/CFT) with strongly interacting systems
described at a fundamental level by quantum chromodynamics (QCD). Also
called the gauge-gravity correspondence, it allows extraction of information
about four-dimensional strongly coupled gauge theories by mapping them onto
gravitational theories in five dimensions where, because of the weak
coupling, they may be solved much more easily. In the semiclassical
approximation, the QCD generating functional of the quantum field theory is
given by the minimum of the classical action of the 5D theory at the 4-D
asymptotic border of the 5-D space. Thus, in principle one can compute
physical observables in a strongly coupled gauge theory in terms of a weakly
coupled classical gravity theory, which encodes information of the boundary
theory. In the so-called bottom-up approach, a 5D holographic dual to QCD is
constructed and quantitative predictions for soft hadronic quantities are
deduced. A current state of the field and references can be found in the
review by Brodsky. et al. \cite{Brodsky} in which the authors also discuss
an interesting connection between light-front dynamics, its holographic
mapping to gravity in a higher-dimensional anti-de Sitter (AdS) space, and
conformal quantum mechanics. This approach sheds additional light on the
confinement dynamics in QCD in the limit of massless quarks. A different set
of topics can be found in the review by Kim and Yi \cite{Kim}. \bigskip

Nevertheless, since the real holographic dual to QCD is unknown, one must
fall back upon physically motivated prescriptions for the dual theory. This
suggests that naive attempts to model hadronic quantities will inevitably
fail at some level. Indeed, although ADS/CFT models work reasonably well for
hadron masses and form factors encountered in weak and electromagnetic
processes, these successes do not extend to quantities such as hadronic
polarizabilities. The simplest of these are $\alpha _{E}$ and $\beta _{M}$,
which are respectively the electric and magnetic dipole polarizabilities of
the charged pions. These quantities characterize the induced dipole moments
of the pion during $\gamma \pi $ Compton scattering \cite{Holstein}, \cite%
{Gasser}. The moments are induced via the interaction of the photon's
electromagnetic field with the quark substructure of the pion. The incident
photon can be thought of as creating the polarizing fields and the outgoing
photon carries information of the extent to which the hadron has been
polarized. $\alpha _{E}$ is the proportionality constant between the $\gamma
^{\prime }$s electric field and the induced electric dipole moment, while $%
\beta _{M}$ is similarly related to the $\gamma ^{\prime }$s magnetic field
and the induced magnetic dipole moment. A more point-like and strongly bound
system will be less polarizable than an extended, weakly bound system. As
such the polarizabilities are fundamental hadronic characteristics and in
principle computable using the many approximate methods used to solve QCD.
This should also be true for ADS/CFT but a calculation of pionic
polarizabilities by Marquet, Roiesnel, and Wallon \cite{Marquet} in an
arbitrary background dilaton field yielded exactly zero for both quantities.
This is in contradiction with the observed experimental values \cite{Gasser}%
. An earlier calculation by Gao and Xiao \cite{Gao} in an ADS/CFT hard-wall
model also yielded zero (although these authors do not explicitly mention
polarizability).

\bigskip

At one level this is surprising. Naively, a minimal model that couples
photons to structureless spin-zero fields should not yield any structure
information. However one quickly realizes that the fields in the bulk are
allowed to oscillate in all possible ways subject to boundary conditions,
and this endows the "shadow" in 4-D with structure. Earlier, in the work of
Grigoryan and Radyushkin \cite{Grigoryan}, the vertex was shown to be
dressed in a manner that resembles a generalized vector meson dominance
model. These efforts result in electromagnetic form factors of hadrons that
are in fair to good agreement with experiment. So, for example, a reasonable
shape of the pion form factor means that the model correctly reflects an
important aspect of the pion's quark substructure. One can go further: as
shown by de Teramond et al \cite{deT} exploring 3-point functions in a
different kinematic regime leads to generalized parton distributions (GPD's)
that lie within the range of acceptable GPD parameterizations that are
consistent with the data. Nevertheless, this minimal model, whether or not
VMD improved, can go only so far. Zero polarizability falsely suggests that
pion is an elementary particle; obviously its compositeness is inadequately
included in the canonical ADS/CFT approach. \bigskip 

The goal of the present communication is to incorporate hadronic
polarizabilities into ADS/CFT by suggesting additions to the basic QCD
gravity dual. This will involve the introduction of effective vertices
consistent with Lorentz and discrete symmetries. In principle the vertices
are computable in approximations to QCD \cite{Korchin}. We shall permit
vector and axial vector mesons to propagate as intermediate states. As the
simplest non-trivial possibility, consider the Compton scattering of purely
real photons from a charged pion. Of course, more information could be
gained using virtual photons since these allow for access to what are known
as generalized polarizabilities \cite{Guichon}, \cite{Scherer}. The
extraction of $\alpha _{E}$ and $\beta _{M}$, as well as higher multipole
probabilities, depends crucially on using the Low-Gellmann-Goldberger low
energy theorem (LET)\cite{LET}. In principle, any model respecting the
symmetries entering the derivation of the LET should reproduce the
constraints of the LET. It is only terms of second order which contain new
information on the structure of the nucleon specific to Compton scattering.
For a spinless target, with $\omega ,\omega ^{\prime }$ being the incident
and scattered photon energies respectively in the laboratory frame, the
theorem gives the form of the scattering amplitude, 
\begin{equation}
f=-\frac{e^{2}}{m}\mathbf{\varepsilon }^{\prime \ast }\cdot \mathbf{%
\varepsilon }+\alpha _{E}\omega \omega ^{\prime }\mathbf{\varepsilon }%
^{\prime \ast }\cdot \mathbf{\varepsilon }+\beta _{M}\omega \omega ^{\prime
}(\mathbf{\varepsilon }^{\prime \ast }\times \mathbf{\hat{q}}^{\prime
})\cdot (\mathbf{\varepsilon }\times \mathbf{\hat{q}})+O(\omega ^{3})
\label{scatt}
\end{equation}%
This LET needs to be put into a Lorentz invariant form, which calls for
identifying a tensor basis with coefficients free from kinematical
singularities. There have been several careful discussions of this in the
literature \cite{Scherer},\cite{Fearing}. An acceptable set that is model
independent must be based only on the requirement of gauge invariance,
Lorentz invariance, crossing symmetry, and the discrete symmetries. The
simplest and most illuminating basis is that of L'vov et al. \cite{Scherer}
who identify a tensor basis with appropriate Lorentz-invariant amplitudes
that are free from kinematical singularities. A gauge invariant separation
is then made into a generalized Born term containing ground-state properties
only and a residual contribution describing the model-dependent internal
structure. More specifically, these authors show that for a spinless hadron,
the non-Born terms from which polarizabilities can be extracted must lead to
a real photon Compton scattering amplitude of the form, 
\begin{eqnarray}
T &=&T_{Born}+\frac{1}{2}b_{1}(0)f^{\mu \nu }f_{\mu \nu }^{^{\prime
}}+b_{2}(0)P_{\mu }f^{\mu \nu }P^{\rho }f_{\rho \nu }^{^{\prime }},
\label{amp} \\
P_{\mu } &=&p_{\mu }+p_{\mu }^{\prime }, \\
f_{\mu \nu } &=&-i(q_{\mu }\varepsilon _{\nu }-q_{\nu }\varepsilon _{\mu }),
\\
f_{\mu \nu }^{\prime } &=&i(q_{\mu }^{\prime }\varepsilon _{\nu }^{\prime
\ast }-q_{\nu }^{\prime }\varepsilon _{\mu }^{\prime \ast }).
\end{eqnarray}%
In the above, $b_{1}(0)$ and $b_{2}(0)$ are the $q^{2}\rightarrow 0$\ limits
of scalar functions of kinematical invariants and are related to the static
polarizabilities $\alpha _{E}$ and $\beta _{M}$ in the small\ $\omega $\
limit (as measured in the lab frame), 
\begin{eqnarray}
\alpha _{E} &=&-\frac{1}{2m}b_{1}(0)-\frac{m}{2}b_{2}(0)  \label{EP} \\
\beta _{M} &=&\frac{1}{2m}b_{1}(0)  \label{MP}
\end{eqnarray}

\bigskip We now turn towards the ADS/CFT calculation of Compton scattering.
With the notations and conventions as used by Marquet et al.\cite{Marquet},
the bulk action with a scalar field $\Phi $ minimally coupled to the field $%
A_{m}$ is, 
\begin{eqnarray}
S_{0} &=&\int d^{4}xdz\sqrt{-g}\left( -\frac{1}{4}F_{mn}F^{mn}+e^{-\chi
}D^{m}\Phi ^{\ast }D_{m}\Phi +e^{-\chi }\mu _{S}^{2}\Phi ^{\ast }\Phi
\right) ,  \label{S0} \\
D_{m}\Phi &=&\partial _{m}\Phi -ieA_{m}\Phi .
\end{eqnarray}%
The dilaton field\ $\chi (z)$\ breaks conformal symmetry, which is necessary
for the 4-D particles to have non-zero mass.\ A popular choice is $\chi
(z)=k^{2}z^{2}.$ However we shall not commit to any particular choice in
this paper. The 5-D field $A_{m}(x,z)$ is dual to the 4-D electromagnetic
field $A_{\mu }(x)$. Since we are concerned here only with real photons, it
is best to choose $A_{z}=0.$ This forces the condition, $\partial ^{\mu
}A_{\mu }=0.$ There is no dilatonic cutoff on the real photon; it may travel
freely in the bulk. As per usual convention, the latin indices $%
m,n=0,1,2,3,z $ and the greek indices $\mu ,\nu =0,1,2,3.$ The Minkowski
space metric used here is $\eta _{\mu \nu }=(-1,1,1,1)$ and $%
g_{mn}dx^{m}dx^{n}=(R^{2}/z^{2})(\eta _{\mu \nu }dx^{\mu }dx^{\nu }+dz^{2}).$

\bigskip The strategy of going from the action to the scattering amplitude
is well known and will not be repeated here in detail. The main steps are:
a)Find the classical equations of motion and then perform the mode
expansion, subject to defined values of the field on the 4-D asymptotic
border of the 5-D space; b)Re-express the action in terms of the solutions
thus found by expanding to the required order in the electric charge $e$;
c)Functionally differentiate with respect to the boundary values of the
field after using the GKP-Witten relation \cite{GKP},\cite{Witten} $%
Z_{QFT}=Z_{Bulk}.$ In calculating the four-point correlation function one
needs the bulk-to-bulk scalar Greens' function which, after Fourier
transformation, has the mode expansion \cite{Marquet}, 
\begin{equation}
\hat{G}(z,z^{\prime },k)=-\sum\limits_{n=0}^{\infty }\frac{\Phi _{n}^{\ast
}(z)\Phi _{n}(z^{\prime })}{k^{2}+m_{n}^{2}-i\varepsilon }.  \label{prop}
\end{equation}%
The eigenfunctions $\Phi _{n}(z)$ are normalized according to, 
\begin{equation}
\int\limits_{0}^{\infty }dz\frac{e^{-\chi (z)}}{z^{3}}\Phi _{n}^{\ast
}(z)\Phi _{m}(z)=\delta _{mn}.  \label{norm}
\end{equation}%
Marquet et al. \cite{Marquet} give the explicit (and somewhat complicated
form) of the scattering amplitude for the general case where both photons
are virtual. We can easily check their result for the simpler situation
where both photons are real. In that case close to the threshold only the $%
n=0$ term in Eq.\ref{prop} contributes. After using the normalization Eq.\ref%
{norm}, up to $O(\omega ^{2})$ one sees that only the Born contribution
remains, 
\begin{eqnarray}
T &=&(2\pi )^{4}\delta ^{4}(p+q-p^{\prime }-q^{\prime })\mathcal{M}_{Born},
\\
\mathcal{M}_{Born} &=&e^{2}\varepsilon ^{\prime \ast \mu }\left( 2\eta _{\mu
\nu }-\frac{(2p+q)_{\mu }(2p^{\prime }+q^{\prime })_{\nu }}{s+m^{2}}-\frac{%
(2p^{\prime }-q)_{\mu }(2p-q^{\prime })_{\nu }}{u+m^{2}}\right) \varepsilon
^{\nu }.  \label{MB}
\end{eqnarray}%
Thus the coefficients $b_{1}(0),b_{2}(0)$ are identically zero here; the
action in Eq.\ref{S0} gives zero polarizabilities in contradiction with both
expectations and measurements. A hint towards the remedy come from current
algebra \cite{Holstein}. The Compton scattering amplitude can be related via
current algebra/PCAC to that for radiative charged pion decay and involves
both axial and vector currents. \bigskip

The advantages and disadvantages of the ADS/CFT approach to hadron structure become evident here. Implicitly ADS/CFT relies on a large-$N_{c} $ approach, which means all calculation in the bulk are tree level and hence easily performed. However, this means that tensor structures that could potentially be generated by loops are strongly suppressed. A way out is to take a more phenomenological approach that anticipates the correct symmetries for the scattering amplitudes. To this end we supplement Eq.\ref{S0} with an action for charged axial vector fields,
\begin{eqnarray}
S_{a} &=&-\int d^{4}xdz\sqrt{-g}e^{-\chi }\left( \frac{1}{2}a_{mn}^{\ast
}a^{mn}+\mu _{A}^{2}a_{m}^{\ast }a^{m}+\frac{1}{2}eg_{A}F^{mn}(a_{mn}^{\ast
}\Phi +a_{mn}\Phi ^{\ast })\right) ,  \label{SA} \\
a_{mn} &=&\partial _{m}a_{n}-\partial _{n}a_{m}.
\end{eqnarray}%
At this point we note that there is another possible effective vertex $%
F^{mn}a_{m}^{\ast }\partial _{n}\Phi $ which is closely similar to $%
F^{mn}a_{mn}^{\ast }\Phi $ in Eq.\ref{SA}. They are not identical in
general, but close to threshold the two vertices yield exactly the same
scattering amplitudes and hence will not be considered separately. The gauge 
$a_{5}(x,z)=0$ is the most convenient choice. In the ADS/CFT correspondence,
the field $a_{\mu }(x,z)$ is sourced by the axial current $A_{\mu }(x)=\bar{q%
}\gamma _{5}\gamma _{\mu }q$ on the boundary at $z=0$. The classical
equation of motion is obtained from Eq.\ref{SA} and it is easy to see that
the Green's function $\hat{G}_{\mu \nu }(x,z,x^{\prime },z^{\prime })$ obeys,%
\begin{equation}
\left[ \frac{z}{e^{-\chi }}\partial _{z}\left( \frac{e^{-\chi }}{z}\partial
_{z}\right) +\eta ^{\mu \nu }\partial _{\mu }\partial _{\nu }-\left( \frac{%
\mu _{A}^{2}R^{2}}{z^{2}}\right) \right] \hat{G}_{\mu \nu }=\frac{g_{\mu \nu
}}{\sqrt{-g}e^{-\chi }}\delta ^{4}(x-x^{\prime })\delta (z-z^{\prime }).
\end{equation}%
Translational invariance on the $z=0$ boundary allows for a Fourier
transformation,%
\begin{gather}
\hat{G}_{\mu \nu }(x,z,x^{\prime },z^{\prime })=g_{\mu \nu }(z^{\prime
})\int \frac{d^{4}k}{(2\pi )^{4}}e^{ik\cdot (x-x^{\prime
})}h_{A}(z,z^{\prime },k), \\
\left[ \frac{z}{e^{-\chi }}\partial _{z}\left( \frac{e^{-\chi }}{z}\partial
_{z}\right) -k^{2}-\left( \frac{\mu _{A}^{2}R^{2}}{z^{2}}\right) \right]
h_{A}(z,z^{\prime },k)=\frac{1}{\sqrt{-g}e^{-\chi }}\delta (z-z^{\prime }).
\end{gather}%
The spectral decomposition of the axial propagator $h_{A}(z,z^{\prime },k)$
is easily found, 
\begin{equation}
h_{A}(z,z^{\prime },k)=-\sum\limits_{n=0}^{\infty }\frac{\Psi _{An}^{\ast
}(z)\Psi _{An}(z^{\prime })}{k^{2}+M_{An}^{2}-i\varepsilon },  \label{Vprop}
\end{equation}%
where $\Psi _{n}$ obeys, 
\begin{eqnarray}
H_{A}\Psi _{n} &=&-M_{An}^{2}\Psi _{n}, \\
H_{A} &=&\frac{z}{e^{-\chi }}\partial _{z}\left( \frac{e^{-\chi }}{z}%
\partial _{z}\right) -\left( \frac{\mu _{A}^{2}R^{2}}{z^{2}}\right) .
\end{eqnarray}%
\ The normalization is,%
\begin{equation}
\int\limits_{0}^{\infty }dz\frac{e^{-\chi (z)}}{z}\Psi _{An}^{\ast }(z)\Psi
_{Am}(z)=\delta _{mn}.
\end{equation}

For a given source term $K_{\nu }(x,z)$ the axial vector field, up to a free
solution, is obtained from, 
\begin{equation}
a_{\mu }(x,z)=\int d^{4}x^{\prime }dz^{\prime }\sqrt{-g(z^{\prime })}%
e^{-\chi (z^{\prime })}\hat{G}_{\mu \nu }(x,z,x^{\prime },z^{\prime })K^{\nu
}(x^{\prime },z^{\prime }).  \label{axf}
\end{equation}%
For the effective coupling in Eq.\ref{SA} the relevant source term for $%
a_{\mu }(x,z)$\ is, 
\begin{equation}
K_{\nu }(x,z)=eg_{A}F_{\nu \alpha }\eta ^{\alpha \beta }\partial _{\beta
}\Phi .
\end{equation}%
We can now insert Eq.\ref{axf} into the action in Eq.\ref{SA} to get the $%
O(e^{2})$ term, pick out the Fourier components, and then differentiate with
respect to the sources on the 4-D boundary. Only terms up to second order in
the photon energy are kept. After a straightforward calculation one obtains
the following contribution to the scattering amplitude,%
\begin{equation}
\mathcal{M}_{A}=\frac{1}{4}e^{2}g_{A}^{2}P_{\mu }f^{\mu \nu }(q)P^{\rho
}f_{\rho \nu }(q^{\prime })\int \frac{dz}{z^{5}}\frac{dz^{\prime }}{%
z^{\prime 5}}e^{-\chi (z)}e^{-\chi (z^{\prime })}(z^{4}+z^{\prime 4})\Phi
_{0}(z)h_{A}(z,z^{\prime },k)\Phi _{0}^{\ast }(z^{\prime }).  \label{MA1}
\end{equation}%
The function $\Phi _{0}(z)$ is the pion wavefunction which was encountered
earlier in the $n=0$ term in Eq.\ref{prop}.\ \ In Eq.\ref{MA1} the
propagator is evaluated at $k=p+q=p^{\prime }+q^{\prime }$, corresponding to
the s channel Feynman diagram. There is also a crossed channel contribution
with $k=p-q^{\prime }=p^{\prime }-q$ and with photon polarization vectors
exchanged. At the quadratic level, the crossed contribution is equal to that
in Eq.\ref{MA1}. Again, at the quadratic level, they are both of the form
given in the second term of Eq.\ref{amp}. Since $h_{A}(z,z^{\prime
},k)=h_{A}(z^{\prime },z,k),$ the $z^{4}$and $z^{\prime 4}$ contribute
equally in Eq.\ref{MA1}. Finally, using the expression for the Green's
function in Eq.\ref{Vprop} with $k^{2}=-m^{2}$\ \ gives $b_{2}(0)$,%
\begin{eqnarray}
b_{2}(0) &=&e^{2}g_{A}^{2}\sum\limits_{n=0}^{\infty }\frac{C_{n}D_{n}}{%
-m_{0}^{2}+M_{An}^{2}},  \label{MA} \\
C_{n} &=&\int_{0}^{\infty }\frac{dz}{z}e^{-\chi (z)}\Phi _{0}(z)\Psi
_{An}^{\ast }(z),  \label{AN} \\
D_{n} &=&\int_{0}^{\infty }\frac{dz}{z^{5}}e^{-\chi (z)}\Psi _{An}(z)\Phi
_{0}^{\ast }(z).  \label{BN}
\end{eqnarray}%
The sum in Eq.\ref{MA} is expected to converge rapidly with $n$, i.e. as the
number of wavefunction nodes increases. From the EOM close to the UV
boundary where $\chi (z)\rightarrow 0,$\ one finds that the scalar
wavefunctions asymptotically behave as $\Phi _{n}(z)$\ $\sim z^{a}$ with $a=%
\frac{1}{2}+\sqrt{4+\mu _{S}^{2}R^{2}}$. Correspondingly, the axial vector
wavefunctions have \ $\Psi _{n}(z)\sim z^{b}$\ with\ $b=\frac{1}{2}+\sqrt{%
1+\mu _{A}^{2}R^{2}}$. Hence the integrals in Eq.\ref{BN} converge at the
lower limit for $\mu _{S}^{2}>0$ \ or $\mu _{A}^{2}>0.$ If $\mu _{S}^{2}=$ $%
\mu _{A}^{2}=0$ then $B_{n}$ is logarithmically divergent and can be
regulated by putting a cutoff at $z=\varepsilon .$ \bigskip

As yet we do not have a contribution to $b_{1}{(0)}$ as in Eq.\ref{MP}. An
effective interaction reflecting QCD symmetries that rectifies the problem
calls for the exchange of a charged vector meson. Consider therefore a
further supplement to the action that preserves the usual continuous and
discrete symmetries,%
\begin{eqnarray}
S_{\text{v}} &=&-\int d^{4}xdz\sqrt{-g}e^{-\chi }\left( \frac{1}{2}\text{v}%
_{mn}^{\ast }\text{v}^{mn}+\mu _{\text{v}}^{2}\text{v}_{m}^{\ast }\text{v}%
^{m}+\frac{1}{4}eg_{\text{v}}\tilde{F}^{mn}(\text{v}_{mn}^{\ast }\Phi +\text{%
v}_{mn}\Phi ^{\ast })\right) ,  \label{SV} \\
\text{v}_{mn} &=&\partial _{m}\text{v}_{n}-\partial _{n}\text{v}_{m},\text{
\ \ }\tilde{F}^{mn}=\varepsilon ^{mnpqz}F_{pq}.
\end{eqnarray}%
Both $S_{\text{a}},S_{\text{v}}$ have only U(1) symmetry. The $A_{1}$ and $%
\rho $ have very different masses,\ in fact $m_{A_{1}}^{2}\approx 2m_{\rho
}^{2}.$ Thus, we expect $\mu _{\text{v}}^{2}$ (Eq.\ref{SV}) and $\mu _{\text{%
A}}^{2}$ (Eq.\ref{SA}) to have quite different values. Let us note that if
chiral SU(2) $\times $ SU(2) had been an exact symmetry of the ordinary
sort, the $\rho $ meson would have been exactly degenerate with the $A_{1}.$
But, in the model of Erlich et al. an additional condensate field can be
introduced and this forces the masses to be different. On the other hand, in
light front holographic QCD with zero mass quarks, the pion is massless but $%
m_{A_{1}}-m_{\rho }$\ owes to the different values of light-front orbital
angular momentum $L$ for the $A_{1}$\ and $\rho $ (see Eq.5.9 of Brodsky
et al. \cite{Brodsky})

\bigskip

The leading order scattering amplitude coming from the vector meson part is, 
\begin{equation}
\mathcal{M}_{V}=e^{2}g_{A}^{2}m^{2}\tilde{f}^{\mu \nu }(q)\tilde{f}_{\mu \nu
}(q^{\prime })\int \frac{dz}{z^{5}}\frac{dz^{\prime }}{z^{\prime 5}}e^{-\chi
(z)}e^{-\chi (z^{\prime })}(z^{4}+z^{\prime 4})\Phi _{0}(z)h_{V}(z,z^{\prime
},k)\Phi _{0}^{\ast }(z^{\prime }).
\end{equation}%
Using $\tilde{f}^{\mu \nu }(q)\tilde{f}_{\mu \nu }(q^{\prime })=-4f^{\mu \nu
}(q)f_{\mu \nu }(q^{\prime })$, and adding in the contribution from the
crossed channel, we then compare with Eq.\ref{MP} to obtain,%
\begin{eqnarray}
b_{1}(0) &=&-e^{2}m^{2}g_{V}^{2}\sum\limits_{n=0}^{\infty }\frac{E_{n}F_{n}}{%
-m_{0}^{2}+M_{Vn}^{2}},  \label{b1} \\
E_{n} &=&\int_{0}^{\infty }\frac{dz}{z}e^{-\chi (z)}\Phi _{0}(z)\Psi
_{Vn}^{\ast }(z),  \label{bc} \\
F_{n} &=&\int_{0}^{\infty }\frac{dz}{z^{5}}e^{-\chi (z)}\Psi _{Vn}(z)\Phi
_{0}^{\ast }(z).  \label{bd}
\end{eqnarray}

The couplings $g_{A},g_{V}$ are determined by the charged meson decay
amplitudes for the axial vector and vector mesons respectively. The lowest
lying axial-vector meson is the $A_{1}$ and the corresponding vector meson
is the $\rho $. From Eq.\ref{SV}, after using translational invariance on
the 4-D boundary, one immediately sees that, 
\begin{eqnarray*}
\mathcal{M}_{A_{1}\rightarrow \gamma \pi } &=&(2\pi )^{4}\delta
^{4}(k-q-p)eg_{A}(\varepsilon ^{^{\prime }\ast }\cdot \varepsilon q\cdot
k-\varepsilon \cdot q\varepsilon ^{^{\prime }\ast }\cdot k)C_{0} \\
\mathcal{M}_{\rho \rightarrow \gamma \pi } &=&(2\pi )^{4}\delta
^{4}(k-q-p)eg_{V}\varepsilon ^{\mu \nu \rho \sigma }k_{\mu }\varepsilon
_{\nu }q_{\rho }\varepsilon _{\sigma }^{^{\prime }\ast }E_{0}
\end{eqnarray*}%
$C_{0},E_{0}$ are the overlaps in Eqs.\ref{AN},\ref{bc}. Knowing the
electromagnetic decay widths of the $A_{1}$ and $\rho $ one can find $%
g_{A},g_{V}$ for any dilaton profile and choice of parameters $\mu _{\text{S}%
},\mu _{\text{A}},$ $\mu _{\text{V}}$ and hence the electric and magnetic
polarizabilities from Eqs. \ref{MA},\ref{b1} together with Eqs. \ref{EP},\ref%
{MP}. The widths $\Gamma _{A_{1}\rightarrow \gamma \pi }$ and $\Gamma _{\rho
\rightarrow \gamma \pi }$ have been measured, albeit quite imprecisely, and
the constants $g_{A},g_{V}$ can be related to them. We shall not pursue
numerical possibilities here, having obtained a minimal extension of the
usual ADS/CFT model that can in principle accommodate hadronic
polarizabilities.

\bigskip

\bigskip {\Large Acknowledgments\medskip }

The author thanks Stan Brodsky, Joshua Erlich, and Guy De Teramond for
helpful correspondence and for valuable comments on the first draft of this
paper.

\end{document}